\RequirePackage{fix-cm}

\documentclass[smallextended]{svjour3}       
\smartqed  
\usepackage{graphicx}
\usepackage{amssymb}
\usepackage[hyphens]{url}
\usepackage[normalem]{ulem}
\usepackage{amsmath}
\usepackage{float}
\usepackage{color}
\usepackage{lineno}
\usepackage{nameref,hyperref}

\graphicspath{{./BMB_Images_May2018/}}

\begin{document}

%
%
\title{Modeling the prescription opioid epidemic}

%
%
\author{Nicholas A. Battista  \and Leigh B. Pearcy \and W. Christopher Strickland
}

\institute{N.A. Battista \at
              Dept. of Mathematics and Statistics, The College of New Jersey, 2000 Pennington Road, Ewing, NJ 08628 \\
              \email{battistn@tcnj.edu}           
           \and
           L.B. Pearcy \at
              Dept. of Mathematics, University of North Carolina at Chapel Hill, CB 3250, University of North Carolina, Chapel Hill, NC, 27599 \\
              \email{leighbp@live.unc.edu}
           \and 
           W.C. Strickland\at
              Dept. of Mathematics, University of Tennessee at Knoxville, 1403 Circle Drive, Knoxville TN, 37996 \\
              (865)\ 974-4299\\
              \email{cstric12@utk.edu}\\ 
              \url{https://www.christopherstrickland.info/}
}

%
%
\date{Received: date / Accepted: date}

\maketitle

%
%

\begin{abstract}

Opioid addiction has become a global epidemic and a national health crisis in recent years, with the number of opioid overdose fatalities steadily increasing since the 1990s. In contrast to the dynamics of a typical illicit drug or disease epidemic, opioid addiction has its roots in legal, prescription medication - a fact which greatly increases the exposed population and provides additional drug accessibility for addicts. In this paper, we present a mathematical model for prescription drug addiction and treatment with parameters and validation based on data from the opioid epidemic. Key dynamics considered include addiction through prescription, addiction from illicit sources, and treatment. Through mathematical analysis, we show that no addiction-free equilibrium can exist without stringent control over how opioids are administered and prescribed, effectively transforming the dynamics of the opioid epidemic into those found in a purely illicit drug model. Numerical sensitivity analysis suggests that relatively low states of endemic addiction can be obtained by primarily focusing on medical prevention followed by aggressive treatment of remaining cases - even when the probability of relapse from treatment remains high. Further empirical study focused on understanding the rate of illicit drug dependence versus overdose risk, along with the current and changing rates of opioid prescription and treatment, would shed significant light on optimal control efforts and feasible outcomes for this epidemic and drug epidemics in general.


%
%
\end{abstract}

%
%

%
%

\section{Introduction}
\label{section:intro}

Starting in the mid 1990s, allegations arose that the medical field systematically under-treated pain, and the American Pain Society lobbied to have pain recognized as a $5^{th}$ vital sign which, if adopted, would require all physicians to accept and treat patient pain reports - naturally leading to an increase in opioid prescriptions and increasing profits for drug manufacturers \cite{VanZee:2009,Mandell:2016}. Meanwhile, confounding medical literature appeared suggesting that cancer patients using prescription opioids to treat their chronic pain did not become addicted \cite{Porter:1980,Perry:1982,Schug:1992}. One study found that only $1$ participant out of $550$ developed an addiction to their prescription painkillers \cite{Schug:1992}. Another study found \textit{no} cases of addiction among 10,000 burn victims using prescription opioid drugs \cite{Perry:1982}. With this data, it began to appear as though physicians could safely prescribe opioids to those in chronic pain without fear of addiction.

By 2000, the Joint Commission began requiring that health care organizations assessing and treat pain in all patients \cite{Mandell:2016}. OxyContin prescriptions for non–cancer-related pain increased from $670,000$ in $1997$ to nearly $6.2$ million in $2002$ \cite{VanZee:2009}. This trend continued through the early 2000s, and in 2012, it was discovered that $259$ million opioid prescriptions had been written - enough for every adult in America to have at least one bottle of pills \cite{CDC:2014}. By 2014, almost 2 million Americans abused or were dependent on prescription opioids \cite{NSDUH:2015}.

Unfortunately, the increase in opioid prescriptions has led to an increase in opioid addiction and abuse, affecting all age demographics. Large  quantities of unused prescription drugs are currently available in many prescribed users' homes \cite{Bicket:2017}, and in 2015, $276,000$ American adolescents were abusing painkillers for non-medical reasons \cite{NSDUH:2015} - many of whom obtained them from a friend or relative who had a prescription \cite{Twombly:2008,Han:2017}. In older age groups, regular, long-term opioid use is more common \cite{Campbell:2010} with possibly $1$ in $4$ long-term opioid users struggling with addiction \cite{Boscarino:2010}. 
Geographically, the opioid epidemic not only affects densely populated areas, but hits rural areas especially hard as well \cite{Keyes:2014}.



Misconceptions regarding prescription opioids make them especially dangerous and include the following: (1) Since opioids are medically prescribed they are safe, (2) you cannot get addicted to prescription painkillers if taken as prescribed, (3) a person is able to safely self-medicate for pain with opioids, (4) only long-term use of certain opioids produces addiction \cite{Twombly:2008,Volkow:2016}. The coupling of these misconceptions with the general availability of opioids makes this epidemic unlike previous drug waves. To make matters worse, many opioid addicts switch to heroin as a cheaper alternative to prescription opioids \cite{Muhuri:2013}, with estimates suggesting that as many as $4$ out of $5$ new heroin users had abused prescription painkillers prior to starting heroin \cite{Jones:2013}. This is contrary to previous trends of addiction moving from heroin use to prescription painkillers abuse in the mid-1950s \cite{Hughes:1972,Lankenau:2012}.

As of October 26, 2017, the US Department of Health and Human Services has declared the opioid crisis a public health emergency \cite{Davis:2017}. Yet despite the current seriousness and scale of the opioid epidemic, the need for effective intervention strategies, and an abundance of literature on mathematical epidemiology for infectious diseases, rigorous mathematical theory has yet to be applied to opioid addiction as it has for other diseases. In fact, very little has been published applying mathematical epidemiology to the problem of drug use in general. White and Comiskey \cite{White:2007} published perhaps the first such model, mathematically describing the heroin epidemic as a system of differential equations resembling the classic SIR model of Kermack and McKendrick \cite{Kermack:1927}. Alterations of this model were subsequently studied by several authors including \cite{Nyabadza:2010,Samanta:2011,Huang:2013,Bin:2015,Ma:2017}, all targeting heroin. In 2012, Njagarah and Nyabadza \cite{Njagarah:2012} described a model exploring the dynamics of drug abuse epidemics more generally, focusing on the interplay between light users, heavy users, and rehabilitation. However, to the authors' knowledge, no one has developed a compartmental differential equation model specifically for prescription opioids with the intent of better understanding the dynamics involved.

In this paper, we investigate the dynamics driving the opioid epidemic by formulating and analyzing an SIR-type model \cite{Kermack:1927,May:1979} built specifically to study addiction to a general class of prescription drugs. Our model includes multiple routes leading to dependency and addiction that are specific to prescription medication, including a ``prescribed'' class that both directly feeds the addicted population and contributes secondary cases via unsecured or unused drugs. We then analyze the model for key properties and conditions that may lead to a meaningful reduction in the number of addicted people and discuss our conclusions. The goal of this paper is to investigate broad trends in prescription opioid addiction rather than localized interactions in order to narrow down possible global strategies for arresting the epidemic long-term.

\section{Mathematical Methods}
\label{section:model}

%
%

We begin by defining $4$ population classes:

\begin{enumerate}
    \item S (\textit{``susceptibles''}): This is the general class of individuals who are not using opioids or actively recovering from addiction.
    \item P (\textit{``prescribed users''}): This class is composed of individuals who are prescribed opioids but do not have an addiction to them. Members have some inherent rate ($\alpha$) of becoming addicted to their prescriptions, and a rate of finishing their prescription without addiction ($\epsilon$).
    \item A (\textit{``addicted''}): All addicted opioid users are here, regardless if their drugs are prescribed. There are multiple routes to this class in our model. One is prescription-induced ($\gamma$) addiction, while two others routes bypass P: one based on interactions with addicted users or their dealers ($\beta(1-\xi)$) and another based on interactions with opioid patients ($\beta\xi$), e.g. unsecured or extra drugs \cite{Bicket:2017}. Addicted users enter treatment at rate ($\zeta$). Here, we define an addicted individual as someone exhibiting a pattern of continued nonmedical use with potential for harm \cite{Vowles:2015}. We will assume throughout this paper that the term ``pain reliever use disorder'', which appears regularly in government reports \cite{NSDUH:2015}, satisfies this definition and that persons who ``misuse'' prescription opioids without further explanation do not satisfy the definition.
    \item R (\textit{``rehabilitation/treatment''}): This class contains individuals who are in treatment for their addiction. We include an inherent rate of falling back into addiction ($\sigma$) as well as a mode of relapsing due to general availability of the drug ($\nu$). This in contrast to White and Comiskey \cite{White:2007} who only allow for the latter approach. Also different in our model: members of the recovering class who complete their treatment (at rate $\delta$) return to being susceptible. That is, we assume successful treatment does not impart permanent immunity to addiction.
\end{enumerate}

\begin{figure}[H]
    \centering
    \includegraphics[width=0.75\linewidth]{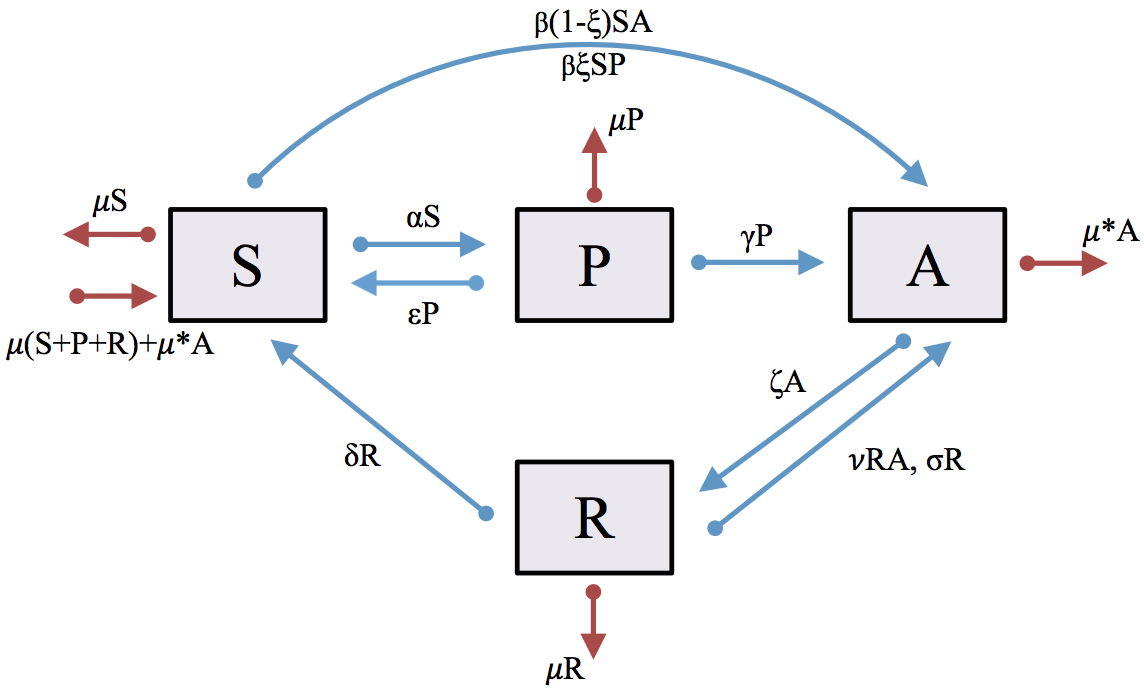}
    \caption{\textbf{Opioid Model Schematic.} A schematic diagram showing the relationships between all the classes in the compartmental model of opioid addiction given by Eqns. \ref{eq:S}-\ref{eq:R}. Red arrows denote death rates, which are passed back into $S$ to maintain a constant population}
    \label{fig:schematic}
\end{figure}

The system is illustrated in Fig. \ref{fig:schematic} and is specified via four continuous-time differential equations
%
%
\begin{align}
    \dot{S} &= - \alpha S - \beta(1-\xi) SA - \beta\xi SP + \epsilon P + \delta R + \mu(P+R)+\mu^{*}A\label{eq:S} \\
    \dot{P} &= \alpha S - (\epsilon+\gamma+\mu) P\label{eq:P}\\
    \dot{A} &= \gamma P + \sigma R +  \beta(1-\xi)SA + \beta\xi SP + \nu RA - (\zeta+\mu^{*})A\label{eq:A}\\
    \dot{R} &= \zeta A - \nu RA- (\delta+\sigma+\mu) R\label{eq:R}.
\end{align}
where $S+P+A+R=1$ and $S,P,A,$ and $R$ represent the mean expected fraction of the population for each class. Time $t$ is understood to be in years, and all rates can be assumed to be yearly rates.

We assume that any mortality due to opioid-related overdose is insufficient to significantly change total population proportions ($S$,$P$,$A$, and $R$), and all deaths are recycled back into the $S$ class to maintain the relation $S+P+A+R=1$. Additionally, we attempt to simply the system by considering only a first-order addiction rate $\gamma P$ from the $P$ class to the $A$ class, assuming that prescribed medication (perhaps from multiple doctors) is the primary source of opioids for most of these users. Second order effects due to mass action with $A$ and other members of $P$ in the $P\rightarrow A$ route would likely also need to include feedback effects including a dynamic whereby large numbers of addicted promote additional caution in the prescribed and susceptible class, and we felt that this study was beyond the scope of a first model for prescription opioid addiction. One exception to this thinking was made for the $\nu RA$ mass-action term, as we wanted to consider possible peer pressures involved in making a successful recovery from addiction. This term represents a balance between the negative influence of prescribed users and the possibility that they would take actions to prevent a recovering addict from accessing their opioids, and so we did not include a $RP$ term. In any case, our results suggest that the model is quite insensitive to the choice of $\nu$ compared to $\sigma$, and so this term would be a good candidate for omission in future studies.

We estimated parameter values from the literature wherever possible with the goal of focusing our attention on a neighborhood of likely values. These estimations are given in Table \ref{tab:params}.

The 2017 CDC Annual Surveillance report states that in 2016, 19.1 out of 100 persons received one or more opioid prescriptions \cite{CDCAnnualSurveillance:2017}. As some of these will have been continuing patients from the previous year, we assume that $\alpha$, our yearly rate of moving from $S$ to $P$, is less than 19.1. We were unable to find more specific data on this rate and so estimated that $\alpha=0.15$. $\epsilon$, the rate of ending opioid prescription use per prescription user-year, was even more difficult to find data on. Most patients end opioid use in less than a month, while a smaller fraction can continue using opioid for over three years \cite{CDCMMWR:2017}. For this reason, we explored a range of values for $\epsilon$ from 0.8 to 8 representing a general belief that most users will quit using prescription opioids in under a year if they have not become addicted.

Our prescription-induced addiction rate ($\gamma=0.00744$) is based off of a comprehensive review \cite{Vowles:2015} which sifted through many opioid patient addiction studies of varying quality and methodology and found significant variance in the addiction rates of prescription opioid users who had been on their prescriptions for at least 90 days (95\% confidence interval would have a rate of approximately 0.057-0.169 in an unweighted collection of studies). Taking only the high quality studies and an average of the minimum and maximum percents, we estimated that 9.3 percent of chronic, non-cancer pain patients become addicted to their opioid prescriptions. Using data that 0.75 of people using prescription opioids for 3 months go on to use for a year and that 0.06 of all initiates to prescription opioids use for a year \cite{CDCMMWR:2017}, we arrived at our value for $\gamma$ as a rate for addictions per prescribed user-year.

We then derived an illicit-induced addiction rate ($\beta=0.0036$) based on the ratio of physician-based sources of prescription opioids to other sources among adults reporting prescription opioid use disorder \cite{Han:2017}, and given that the Substance Abuse and Mental Health Services Administration (SAMHSA) suggests that $2.1$ million people abused prescription opioids for the first time in 2015 out of a population of $320 \mbox{ million}$ \cite{NSDUH:2015}. Using this same source data \cite{Han:2017}, $\beta$ was then further subdivided by differentiating the cases in which a user primarily obtains opioids from friend/relative/other ($\xi\beta$) or from a source related to general addictive demand (drug dealers/strangers) ($(1-\xi)\beta$), where $\xi$ was estimated to be 0.74. These parameter estimates are only meant to be rough starting points for the purpose of basic analysis, particularly as we expect these numbers to vary with both time and location.

The literature broadly suggests that approximately $90\%$ of those entering treatment relapse during the first year in recovery \cite{Smyth:2010,Bailey:2013,Weiss:2017}. While we assume that this rate would be lower if the overall supply and demand of illicit drugs was reduced, it is hard to tease out to what extent. Acute stage withdrawal lasts at most a few weeks \cite{Gossop:1987}, and studies on heroin addicts suggest that up to $70\%$ of recovering addicts may relapse during the first month after treatment ends \cite{Smyth:2010,Bailey:2013}. A study on US prescription opioid addicts (no heroin) similarly found that 8 weeks after cessation of treatment, only 9\% had not relapsed \cite{Weiss:2017}. We could not find published data on 4 weeks post-treatment. Therefore, we took the timescale of recovery and relapse to be approximately one year and made an estimate of 0.7 for the base relapse rate $\sigma$. This estimate is based on the assumption that if there is no relapse within the first weeks, relapse is either due to factors other than primary withdrawal or that there will be no relapse. $\nu$, the relapse rate ascribed to the presence of other addicted individuals, was then assigned a value of 0.2 so that $\sigma+\nu=0.9$, the estimated yearly proportion of $R$ that relapse. While this is certainly a very rough estimate that neglects factors such as post-acute withdrawal and discrepancy in treatment methodology over both time and location, we believe it to be a reasonable enough guess to serve as a gross estimate in this first model.


To estimate $\mu^*$, the overall death rate for prescription opioid addicts, we took a rough estimate of the number of prescription opioid deaths attributed to addicted individuals, $54.6\%$ \cite{Gwira:2014}, multiplied it by the prescription opioid death rate for the entire population \cite{Seth:2018}, and then normalized the result to find the rate of prescription opioid deaths for the addicted class \cite{NSDUH:2015}. We then added the natural death rate $\mu$ into it \cite{Kochanek:2017}, e.g. 
\[
\mu^*= 0.546 \times \left(\frac{5.2}{100,000}\right) \times \left(\frac{300}{2.0 }\right) + \left(\frac{728.8}{100,000}\right) = 0.01155.
\]

%
%
%



 \begin{table}
    \centering
    \caption{\textbf{Table of estimated parameters for the opioid model (all rates are per-capita yearly rates)}}
    \begin{tabular}{llcc}
        \ & Description & Est. Value & Ref. \\ \hline
        $\alpha$ & prescription rate per person per year & 0.15 & \cite{CDCAnnualSurveillance:2017}\\ \hline
        $\epsilon$ & end prescription without addiction (rate) & $0.8-8$ & \cite{CDCMMWR:2017} \\ \hline   
        $\beta$ & total illicit addiction rate for $S$-class & 0.0036 & \cite{Han:2017,NSDUH:2015} \\ \hline
        $\xi$ & fraction of $\beta$ due to $P$ & 0.74 & \cite{Han:2017}     \\ \hline
        $\gamma$ & prescription-induced addiction rate & $0.00744$ & \cite{Vowles:2015,CDCMMWR:2017} \\ \hline
        $\zeta$ & rate of $A$ entry into rehabilitation & 0.2--2 & \\ \hline
        $\delta$ & successful treatment rate & 0.1 & \cite{Weiss:2017} \\ \hline
        $\nu$    & relapse rate of $R$-class due to $A$ & $0.2$ & \cite{Smyth:2010,Bailey:2013} \\ \hline
        $\sigma$ & natural relapse rate of $R$-class & $0.7$ & \cite{Smyth:2010,Bailey:2013}\\ \hline
        $\mu$     & natural death rate & 0.007288 & \cite{Kochanek:2017} \\ \hline
        $\mu^{*}$ & death rate of addicts & 0.01155 & \cite{Gwira:2014,NSDUH:2015,Kochanek:2017,Seth:2018} \\
    \end{tabular}
    \label{tab:params}
\end{table}



%
%
%
%

\section{Results}
\label{section:results}

%
%

%
%

\begin{figure}[ht]
    \centering
    \includegraphics[width=0.70\linewidth]{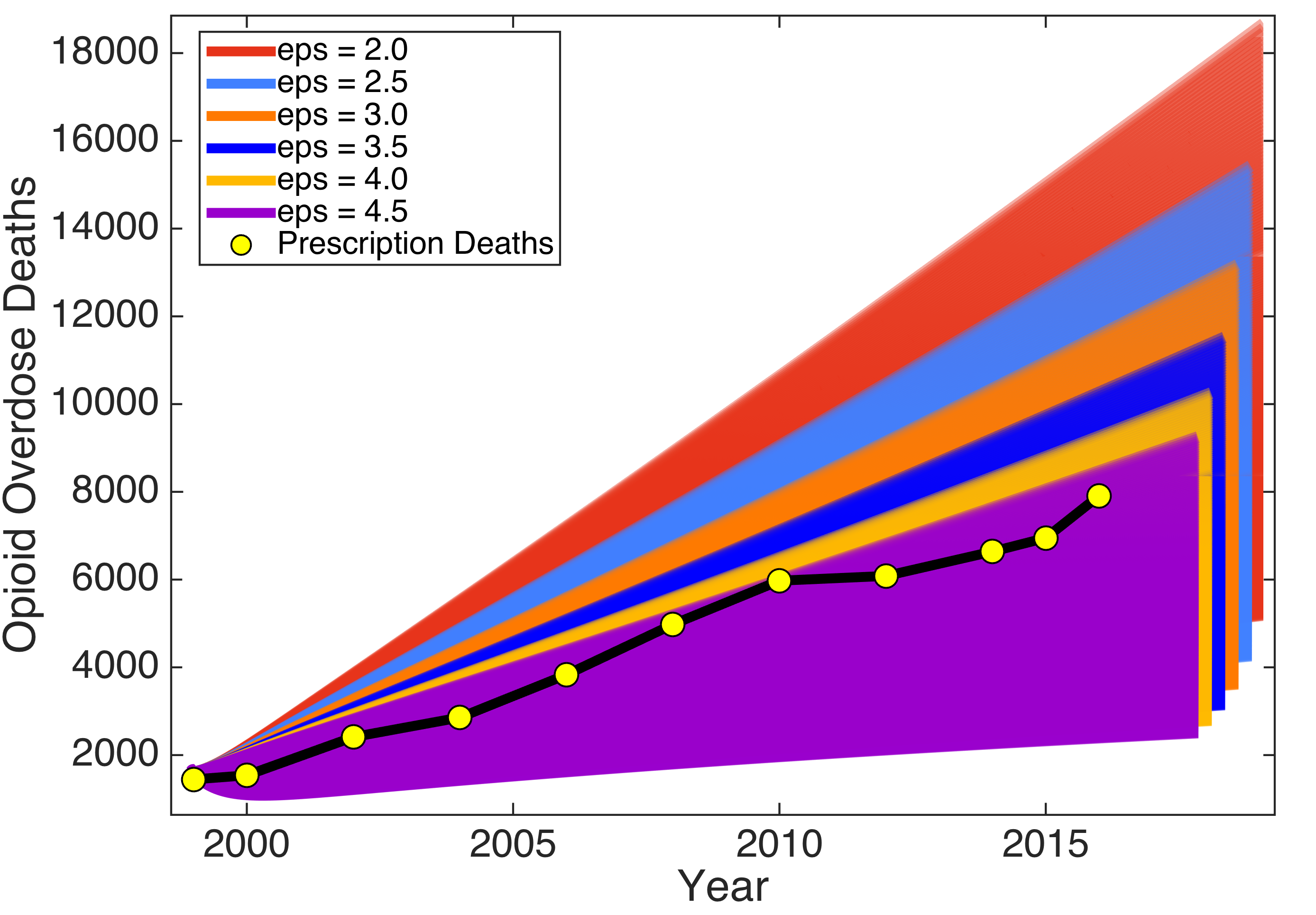}
    \caption{\textbf{Model Validation.} Validating the model for varying $\epsilon\in[1,4]$ and $\zeta\in[0,1]$ against prescription opioid death data from \cite{Hedegaard:2017}}
    \label{fig:validation}
\end{figure}

The model was validated against national data for prescription opioid deaths between 1999 and 2016 \cite{Hedegaard:2017}. To estimate the proportion of these fatalities that could be attributed specifically to addicted individuals rather than misuse by others, we adopted the percentage of prescription opioid deaths ($54.6\%$) attributed to persons who had one or more high-risk factors, such as greater than $4$ prescribers, $4$ different pharmacies, or a daily dosage greater than 100 morphine milligram equivalents (MME) \cite{Gwira:2014}. 
Simulations were then carried out using the estimated parameter values from Table \ref{tab:params} (see Fig. \ref{fig:validation}) and initial conditions chosen to approximate the proportion of each model compartment class present in 1999 (see Appendix \ref{app:initial_conds}). In each simulation, we varied the rates of ending opioid prescriptions without addiction ($\epsilon$) and treatment initiation ($\zeta$). The number of simulated opioid related deaths were then found by computing $pop(t)\times(\mu^*-\mu)A(t)$, where $pop(t)$ was computed by taking the US population between 1999 and 2016 and finding the best fit line through the data \cite{USCensus:2018}.

\begin{figure}[hbtp]
    \centering
    \includegraphics[width=0.70\linewidth]{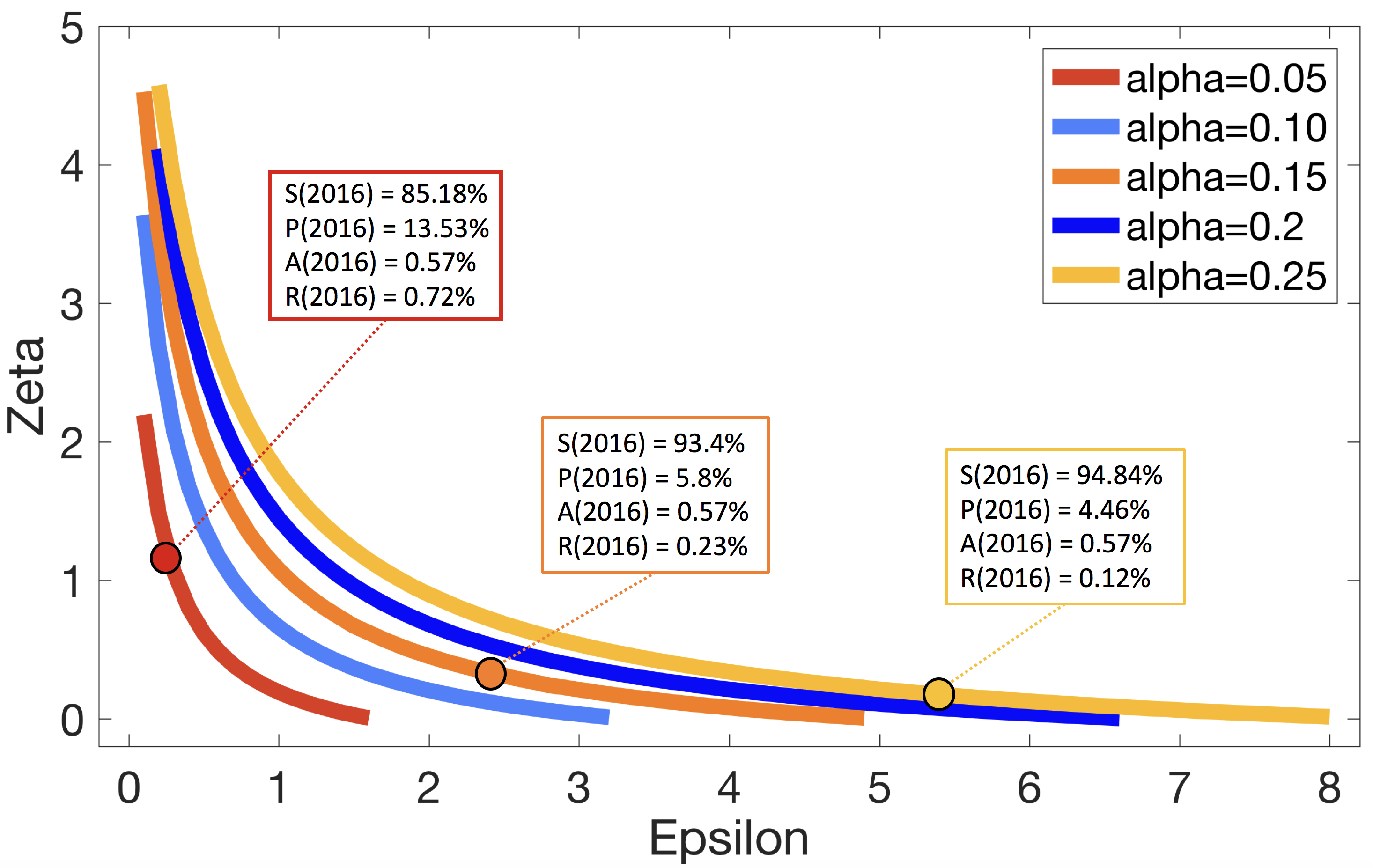}
    \caption{\textbf{Model Validation.} Validating the model for varying $\epsilon\in[1,4]$ and $\zeta\in[0,1]$ against prescription opioid death data from \cite{Hedegaard:2017}}
    \label{fig:validation2}
\end{figure}

Each color in Fig. \ref{fig:validation} corresponds to a particular $\epsilon$ value with $\zeta\in[0,1]$. Our model generally agrees with the data for over a range of $\epsilon$ and $\zeta$ values. 
Additionally, we explored which combinations of $\alpha,\epsilon,$ and $\zeta$ would exactly predict the number of 2016 opioid overdose deaths attributed to individuals who are addicted. These relationships are found in Fig. \ref{fig:validation2}. A few of the points on the feasibility curves were also chosen to highlight population fractions within realistic ranges. For example, when $\alpha=0.05$, $\epsilon=0.30$, and $\zeta=1.069$, we find population fractions of $S(2016)=0.8517$, $P(2016)=0.1353$, $A(2016)=0.0057$, and $R(2016)=0.0072$. Or when $\alpha=0.25$, $\epsilon=5.30$, and $\zeta=0.183$, we predict population fractions of $S(2016)=0.94$, $P(2016)=0.0446$, $A(2016)=0.0057$, and $R(2016)=0.0012$. Roughly 2 million Americans had a substance abuse disorder involving prescription opioids in 2016, hence roughly $2\times 10^6/300\times10^6 \text{ US Pop} = 0.0066$ of Americans were actually addicted to prescription opioids, though this number is for the entire year. Between 1998 and 2006, one estimation for $P$ is that 2\% of adults were taking an opioid in any given week \cite{Boudreau:2009}, so we might expect the actual value for $P$ to be somewhat greater than 0.02 in 2016.

%
%
%

\subsection{Addiction-free equilibrium}

Existence of an addiction-free equilibrium (AFE) is dependent on the condition that $\gamma=0$, and either that $\beta=0$ or $\xi=0$. If $\beta=0$, opioids are no longer available anywhere in our model causing addiction to die out and remain extinct since $\gamma=0$ as well. If $\xi=0$ instead, opioids are available only because of demand from current addicts through the black market - essentially reducing the model to a sub-case that is applicable to any non-prescription illicit drug epidemic. In this case, the AFE is given by
\begin{align}
\begin{split}
    S^\ast &= \frac{\epsilon + \mu}{\alpha + \epsilon + \mu},\hspace{1.5cm} A^\ast = 0,\\
    P^\ast &= \frac{\alpha}{\alpha + \epsilon + \mu},\hspace{1.5cm} R^\ast = 0.
\end{split}\label{eqn:AFE}
\end{align}

%
%

Traditionally, the basic reproduction number denotes how many secondary infections result from one infected individual within a population. When $\mathcal{R}_0>1$, the epidemic is expected to grow as more infections occur while for $\mathcal{R}_0<1$, the number of infected individuals declines. This remains consistent in the context of drug epidemics: $\mathcal{R}_0$ may be calculated and denotes how many addictions there will be in the next generation (year) compared to the current one. Assuming that $\gamma=0$, $\xi=0$ and $\beta> 0$, $\mathcal{R}_0$ can be found using the next generation method \cite{Heffernan:2005,Diekmann:2009}:
\begin{equation}
    \begin{gathered}
    \mathcal{R}_0 = \frac{\beta(\epsilon + \mu)}{(\alpha + \epsilon +\mu)(\mu^\ast+\zeta\Lambda)}=\frac{\beta S^*}{\mu^*+\zeta\Lambda}\\ \text{where}\ \ \ \Lambda=\frac{\delta+\mu}{\delta+\mu+\sigma},\ S^\ast = \frac{\epsilon + \mu}{\alpha + \epsilon + \mu}.\label{eqn:R0_A}
    \end{gathered}
\end{equation}
A derivation of this result is given in Appendix \ref{app:R0}, and is confirmed by Jacobian analysis in Appendix \ref{app:jacobian_analysis}. For parameter values estimated in Table \ref{tab:params}, $\mathcal{R}_0<1$, and so in the absence of prescription-based primary and secondary addictions, we actually expect the opioid epidemic to die out on its own. This result reinforces conventional wisdom that unlike previous drug epidemics, prescription opioid addiction is essentially a by-product of primary and secondary addictions caused by medical prescription and likely would not be self-sustaining absent these prescriptions.

Another potentially surprising result of this calculation is that increasing $\alpha$, the rate at which opioids are prescribed to the general population, actually reduces $\mathcal{R}_0$ and can thus act as a control on the epidemic. This behavior is a direct result of the assumption that $\xi=0$ and $\gamma=0$, which are requirements for the existence of the AFE. If no prescribed users can become addicted to their drugs and their prescriptions do not cause other people to become addicted either (perhaps through tight controls and monitoring), then the prescribed class effectively becomes a safe haven from opioid addiction.

%
%

To analyze the bifurcation of this system with respect to $\beta$ when $\mathcal{R}_0=1$, we follow the method described by Castillo-Chavez and Song \cite{Castillo:2004} and demonstrated by White and Comiskey \cite{White:2007}. Let $\Gamma=\zeta/(\delta+\mu+\sigma)$, and $\Lambda=(\delta+\mu)/(\delta+\mu+\sigma)$. Then whenever
\begin{equation}
\Lambda\Gamma\nu > (1+\Gamma)(\mu^*+\zeta\Lambda),\label{eqn:bcond}
\end{equation}
a backward bifurcation occurs (see \nameref{app:jacobian_analysis} for details of the calculations). Practically speaking, this implies that when Eqn. \ref{eqn:bcond} is satisfied, a positive, stable, endemic equilibrium exists simultaneously with the stable AFE, raising the possibility that additional effort beyond achieving $\mathcal{R}_0<1$ may be required to eliminate addiction. For our estimated parameters in Table~\ref{tab:params}, a backward bifurcation does not occur.

%
%

\subsection{Numerical Sensitivity Analysis}

\begin{figure}[htpb]
    \centering
    \includegraphics[width=\linewidth]{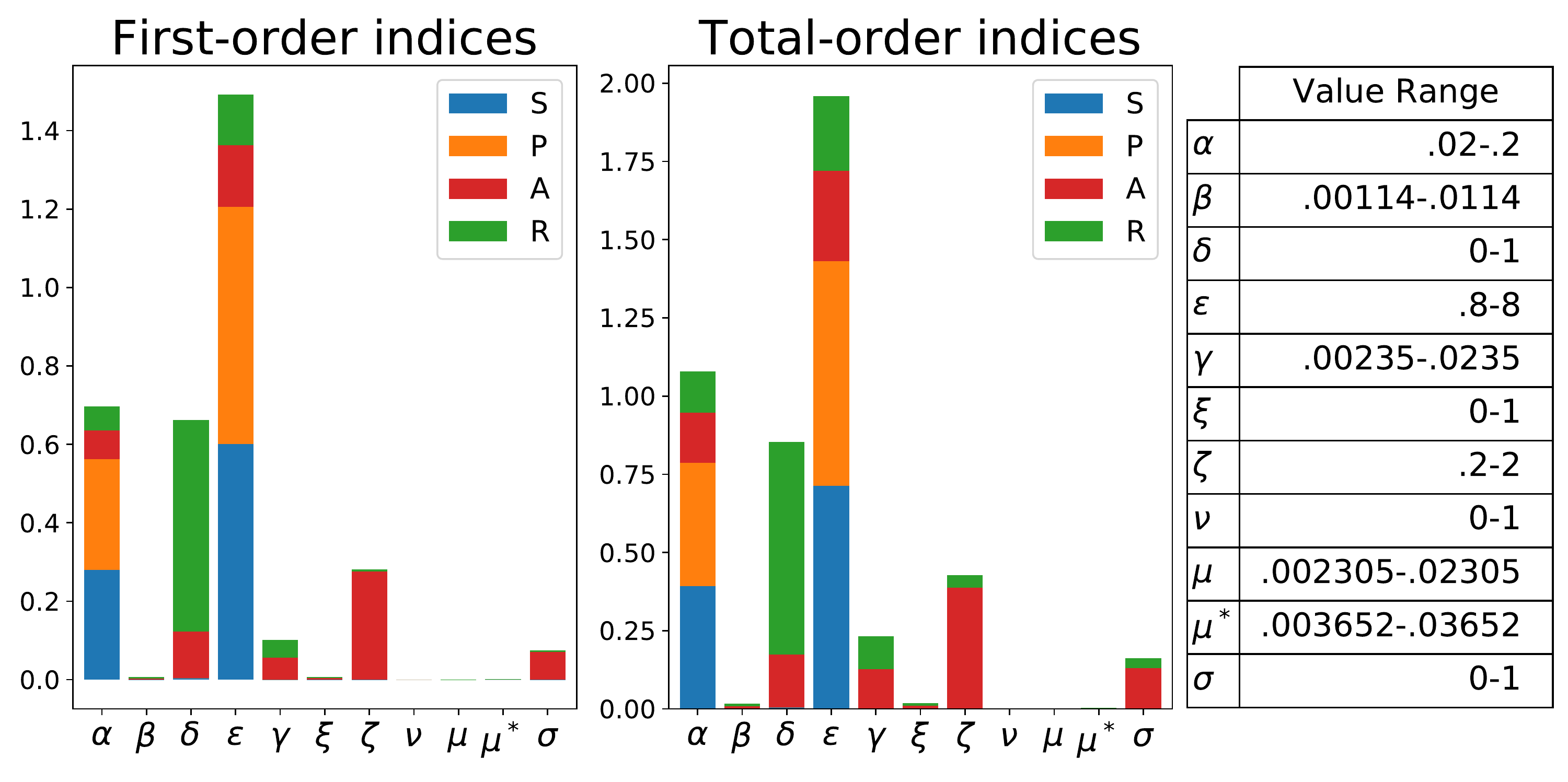}
    \caption{\textbf{Sensitivity of 10 year values for $S,P,A,$ and $R$ to model parameters.} See Fig. \ref{fig:schematic} or Table \ref{tab:params} for parameter definitions. First-order indices do not take into account interactions with other parameters, while total-order indices measure sensitivity through all higher-order interactions}
    \label{fig:10yr_sens}
\end{figure}

To assess the overall 10 year sensitivity of the model to its parameters, we used Saltelli's extension of the Sobol sequence \cite{Saltelli:2002,Saltelli:2010} to vary each parameter within a range about its estimated value. We then conducted Sobol analysis of variance \cite{Sobol:2001} on the resulting values of $S,P,A,$ and $R$ after 10 years. Initial conditions were chosen to reflect estimations of recent U.S. population fractions in each class around the year 2016: $P_0=0.05$ \cite{Boudreau:2009} (some increase added for passage of time), $A_0=0.0062$ \cite{NSDUH:2015}, and $R_0=0.0003$ \cite{TEDS:2016} resulting in $S_0=0.9435$ so that $S+P+A+R=1$. Relative sensitivity of the parameters can be seen in Fig. \ref{fig:10yr_sens}, where longer bars of a given color denote higher sensitivity to that parameter. The reported results for all sub-bars in this figure are within a $95\%$ confidence interval of 0.0056. For parameter sensitivity analysis with respect to the model's AFE, see Appendix \ref{app:AFE_sensitivity}.

%
%
%
%
%
%

\subsection{Simulation Results Around Realistic Parameters}

\begin{figure}[ht]
    \centering
    \includegraphics[width=0.9\linewidth]{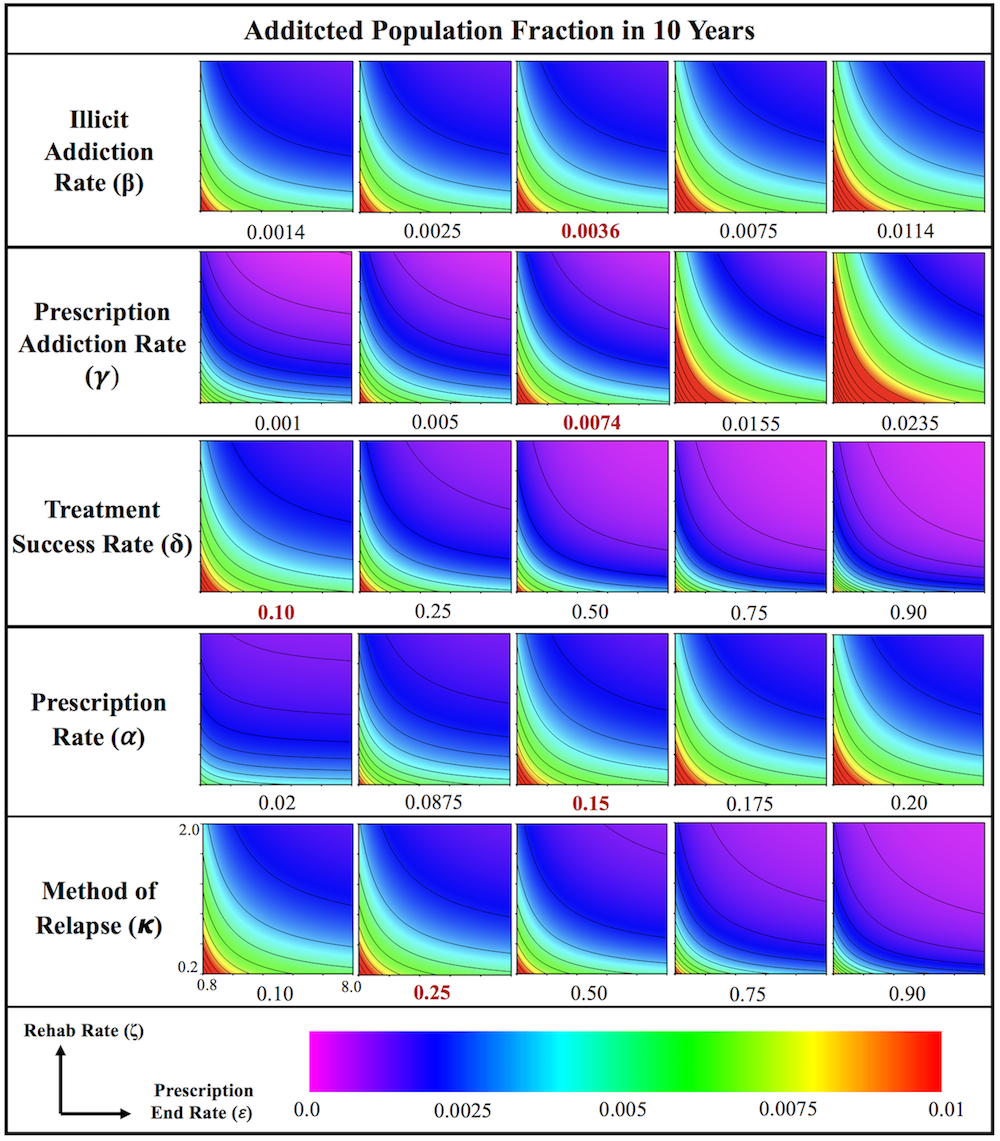}
    \caption{\textbf{Varying the Illicit Addiction Rate ($\beta$), Prescription Addiction Rate ($\gamma$), Treatment Success Rate ($\delta$), Prescription Rate ($\alpha$), and Method of Relapse ($\kappa$).} Colormaps illustrate the predicted 10 years addicted population fraction for prescription completion rates ($\epsilon$) and rehabilitation rates ($\zeta$) between [0.8,8.0] and [0.2,2.0], respectively, while varying the other parameters one at a time}
    \label{fig:combined_results}
\end{figure}

The parameters $\epsilon$ (the rate at which prescribed persons complete their opioid prescription(s)) and $\zeta$ (the rate that addicts enter treatment) are difficult to parse out from data so in the following results, we varied them in the space of $\epsilon,\zeta\in[0.8,8.0]\times[0.2,2.0]$ while simultaneously considering changes in one other parameter at a time: $\beta$, $\delta$, $\alpha$, $\gamma$, and $\kappa$ where $\kappa=\nu/(1-\delta)$ is the ratio of relapse back into addiction attributable to illicit usage ($\nu$) over the total relapse rate ($\delta$ is held constant when changing $\kappa$ and $\nu/(1-\delta) + \sigma/(1-\delta) = \kappa + \sigma/(1-\delta) = 1$, so $\kappa$ gives the fraction of relapse due to $\nu$ vs. $\sigma$). The combined results are shown in Fig. \ref{fig:combined_results}. Whenever unspecified by the plot, all parameters were held constant as in Table~\ref{tab:params} except in the case of $\delta$ and $\kappa$, where it was assumed that $\nu+\sigma+\delta=1$. In the case of $\delta$, it was assumed that $\nu$ and $\sigma$ maintain their original ratio from Table \ref{tab:params} ($\nu=(2/9)(1-\delta)$ and $\sigma=(7/9)(1-\delta)$), and in the case of $\kappa$, $\delta$ was held constant.

%
%

The first row of Fig. \ref{fig:combined_results} examines the effect of varying the overall rate of obtaining illicit opioids ($\beta$) while holding $\xi=0.74$, which dictates the source of the drugs. As suggested by Fig. \ref{fig:10yr_sens}, model results do not appear to be sensitive to $\xi$ and so it was not included in Fig. \ref{fig:combined_results}. To minimize the number of opioid addicts, a high prescription completion rate and a high rate of entering treatment is required. As $\beta$ increases, there exists a higher addicted class for low values of $\epsilon$ and $\zeta$, suggesting that illicit opioids could exacerbate the number of addicted in certain circumstances - a scenario that was not apparent from the Sobol analysis in Fig. \ref{fig:10yr_sens} which was conducted within a larger feasibility space of all parameters rather than the estimated values in Table~\ref{tab:params}. However, as suggested by Fig. \ref{fig:10yr_sens}, the model seems less sensitive to $\beta$ than other parameters explored by Fig. \ref{fig:combined_results}.

On the other hand, the second row of Fig. \ref{fig:combined_results} suggests that for estimated parameters the rate at which medically prescribed opioid users become addicted ($\gamma$) very significantly affects the number of addicted. When $\gamma$ doubles from its assumed realistic value of $0.00744$ to $0.015$, the number of addicts virtually doubles as well. 
For small differences in $\kappa$, the system is not particularly sensitive to how addicts relapse, either on their own ($\sigma$) or based on total illicit usage and availability ($\nu A$), but if $\sigma$ can be made small relative to $\nu$, the number of addicted persons can be expected to fall - an unsurprising result given that $\nu$ appears in a second order term ($\nu RA$) while $\sigma$ appears in a first order term ($\sigma R$). Further analysis, including how each of these parameters affect compartments other than $A$ can be found in the Appendix.

%
%

\begin{figure}
    \centering
    \includegraphics[width=0.925\linewidth]{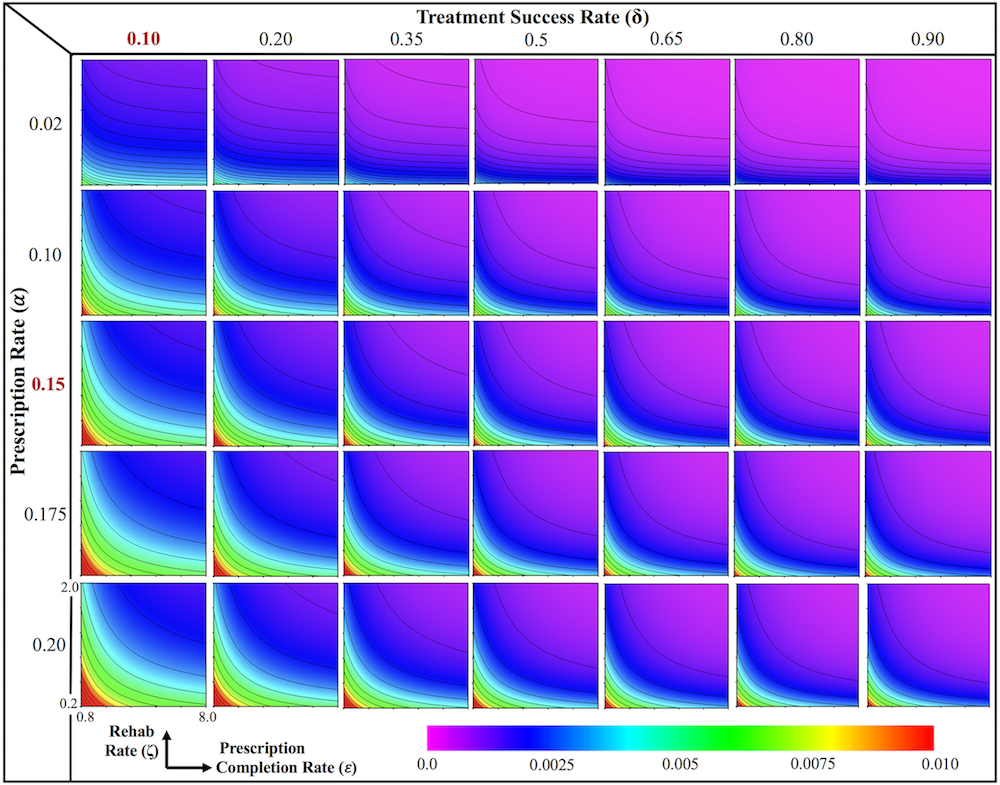}
    \caption{\textbf{Varying the the Prescription Rate ($\alpha$) and Treatment Success Rate ($\delta$)} in tandem with varying prescription completion rates ($\epsilon$) and rehabilitation rates ($\zeta$) between [0.8,8.0] and [0,1]. Colormaps illustrate the predicted 10 years addicted population percentage}
    \label{fig:model_alpha_delta_sweep}
\end{figure}

Finally, we explored the relationship between $\alpha, \epsilon, \delta$, and $\zeta$ in detail, as these parameters are most likely to be the target of control efforts. The results can be seen in Fig. \ref{fig:model_alpha_delta_sweep}.

%
%

\section{Discussion and Conclusion}

In this paper, we present a first model for the opioid epidemic which utilizes the successful mathematical epidemiology approach popularized by Kermack and McKendrick in 1927 for the spread of infectious disease \cite{Kermack:1927}. Parameters are estimated from the literature and simulation results are compared with mortality data and estimates for current population fractions of our given model compartments. Analysis of our model shows that maintenance of an addiction-free population (the addiction-free equilibrium, or AFE) requires at minimum the elimination of both patient prescription-induced addiction ($\gamma=0$) as well as secondary, non-patient addictions attributable to filled prescriptions ($\xi=0$). Parameter sensitivity analysis indicates that the first of these is far more important, with near-AFE endemic states possible even if $\xi$ is significantly greater than zero as long as $\gamma=0$ (see Appendix \ref{fig:gam_xi}). This result strongly suggests that reducing the number of addictions among opioid-prescribed patients is a critical first step in combating the crisis.

Even if in the hypothetical case that both prescription induced addiction and addiction resulting from leftover prescription opioids was eliminated, the threat of ongoing, endemic addiction persists due to illicit availability of these drugs. In this case, our model reduces to a classic illicit drug addiction model except that prescribed opioid users are considered safe from addiction since they are closely monitored to prevent addiction to the drugs they are taking. Our calculation of the basic reproduction number, $\mathcal{R}_0$, then provides a metric by which we can determine if overall addiction will eventually die off or persist based upon model parameters. 

Due to the form discovered for $\mathcal{R}_0$ (Eqn. \ref{eqn:R0_A}), the ratio of addiction to illicit opioids ($\beta$) to the death rate of addicts ($\mu^*$) appears to be critical. If this ratio is less than one, as it is for our parameter estimates, the opioid epidemic is not self-sustaining without prescription drugs no matter the prescription rate or addiction treatment rate. This precise ratio $\beta/\mu^*$ may be somewhat artificial due to the recycling of overdosed persons back into the susceptible class (done in order to maintain an overall static population size). But the suggestion of a natural balance between a drug's infectiousness and potential for addiction verses its potential to be lethal is not far-fetched and could merit further study to better understand addiction in the context of an infectious social disease. 

In parameter regimes where $\mathcal{R}_0\geq1$, we note the importance of entry into rehabilitation ($\zeta$), particularly as the rehabilitation success rate grows ($\delta$) and the natural tendency to relapse shrinks ($\sigma$). This is evidenced both by the parameter sensitivity analysis in Fig. \ref{fig:10yr_sens} which highlights $\zeta$ as a key parameter for the addicted class and Eqn. \ref{eqn:R0_A}, where we see that the effect of $\zeta$ on $\mathcal{R}_0$ is regulated by $\sigma$ and $\delta$.

Given the difficulties of completely eradicating prescription-based addiction ($\gamma=0$ and $\xi=0$), the idea of reaching an addiction-free state remains improbable. Relaxing these assumptions, our results suggest that control efforts should focus on reducing the average prescription length ($\epsilon$) and increasing the rate addicts enter treatment ($\zeta$), even if treatment is often unsuccessful (Fig. \ref{fig:combined_results}), followed by decreasing the number of prescriptions written ($\alpha$). Reducing both $\epsilon$ and $\alpha$ could help naturally decrease the rate of prescription-induced addiction, $\gamma$. In one typical case where $\epsilon=3.0$ and $\zeta=0.25$, doubling the rate that users enter treatment to $\zeta=0.5$, resulted in a $23.5\%$ decrease in the addicted population after 5 years, despite the fact that treatment success was held at $10\%$. If the treatment success rate doubles to $\delta=0.2$, the addicted population decreases another $9.3\%$ (or by $30.5\%$ of where it was initially) after 5 years.


Following this, we found that increasing the success rate of rehabilitation ($\delta$) should also be a priority (Fig. \ref{fig:sims_delta} in the Appendix). It is interesting to note that decreases in the overall length of prescriptions (increasing $\epsilon$) have the most effect when $\epsilon\geq1.5$. The beneficial effect of decreasing overall the prescription lengths is particularly pronounced when either the rate of starting rehabilitation ($\zeta$) is low or the success rate of rehabilitation ($\delta$) is low, regardless of the prescription rate ($\alpha$)
. 
On the other hand, our model suggests that neither the mode of relapse nor the origin of illicit opioids, whether from leftover prescriptions or demand-driven market, has much impact on the total fraction of addicted (Fig. \ref{fig:10yr_sens}, and Figs. and \ref{fig:gam_xi} and \ref{fig:model_relapse_methods} in the Appendix).

To simplify the dynamics for this first model, we neglected potential effects due to gender, race, and geographical location. Additionally, our model did not attempt to capture how prescription drug addicts may move to heroin or vice versa, leaving this study to future work. This dynamic has important ramifications for public health as heroin use is associated with high rates of overdose, especially when laced with fentanyl \cite{Gladden:2016,Peterson:2016,ODonnell:2017}, and could be particularly lethal for users who have first built up an opioid tolerance and then increase their doses on heroin \cite{Muhuri:2013}.
While many have modeled the heroin epidemic previously \cite{Mackintosh:1979,White:2007,Battista:2009,Nyabadza:2010,Huang:2013,Abdurahman:2014}, we are not aware of studies that incorporate effects of fentanyl, methadone, and prescription opioids all together, or studies that explicitly consider demographic effects. Our model is meant to provide a starting point for this larger, more detailed work.

Another simplification we made for the presentation of this first model was the implicit assumption that parameter values are constant with respect to time. This is obviously not the case in for many of our parameters, in particular the prescription initiation rate ($\alpha$) \cite{Pezalla:2017}, the prescription completion rate ($\epsilon$) \cite{Scully:2018}, the rehabilitation initiation rate ($\zeta$), and the rehabilitation success rate ($\delta$). Despite the large amount of public interest in prescription opioid addiction, we found it quite difficult to obtain our ball-park estimates for many of the parameters, as prescription and addiction statistics are often given in yearly aggregate numbers and survey studies are not typically designed with the intent to parameterize mathematical models. For other parameters such as $\beta$, $\xi$, and $\nu$, data is almost wholly absent by nature; fortunately, our results suggest that the model is relatively insensitive to these parameters. While beyond the scope of this particular study, we believe that a rigorous, time-sensitive estimation of model parameters is an important next step and represents a significant work on its own.

In summary, our main results confirm that necessary measures to combating the opioid epidemic include lowering the number and duration of medically prescribed painkillers, more successful treatment regimens, and increasing the availability, ease, and motivation for opioid addicts to enter treatment \cite{Watkins:2017}. Our findings also provide a direct measure of the epidemic's sensitivity to each of these efforts which may be useful in allocating available resources, especially for small rural towns, cities, or states combating the epidemic. Better estimates of model parameters from data could prove crucial in developing management strategies and refining our modeling approaches - given the role of non-prescription opioids such as heroin and fentanyl to the overall epidemic and the unique effects of geography and population demography, we believe that the model presented here represents only the beginning in a larger mathematical exploration of opioid addiction dynamics.


%
%

\begin{acknowledgements}
The authors would like to thank Christina Battista, Robert Booth, Namdi Brandon, Kathleen Carroll, Jana Gevertz, Anne Ho, Shanda Kamien, Grace McLaughlin, Gianni Migliaccio, Matthew Mizuhara, and Laura Miller for comments, suggestions, and informative conversations. NAB would like to thank Patricia Clark of RIT, whose mathematical biology course gave the original motivation for this project in 2009. 
\end{acknowledgements}

%
%
\clearpage
\appendix

\section{Appendix}
%
%


Here we present supplemental material to support our findings including additional model analysis and validation, numerical stability analysis, and simulation data. We also provide details for the calculation determining a condition for backward bifurcation, the explicit Jacobian used in our stability analysis, and simulation results illustrating system sensitivity to the
prescription addiction rate ($\gamma$),
treatment success rate ($\delta$),
prescription rate ($\alpha$), and
method of relapse ($\kappa$). Futhermore, we explore the  and the relationship between prescription rate ($\alpha$) and prescription addiction rate ($\gamma$).

%
%

\subsection{Initial Conditions for Validation}
\label{app:initial_conds}

We estimated the initial prescribed population, $P_0$, based off of the percentage of U.S. population to whom were prescribed opioids at any given week in 2009 ($2\%$) \cite{Boudreau:2009}. Since there were more prescriptions given in 2009 than 1999 \cite{CDCMMWR:2017}, we estimated that roughly $0.40\times 2\%$ of the population were prescribed opioids at any time in 1999, hence $P_0=0.008$. Note we estimated the coefficient of $0.40$ by using the ratio of total opioids MME sold in 1999 to 2009 \cite{FDA:2018}. 

We backed out the initial addicted population from the number of prescription opioid deaths in 1999 (2749) \cite{Hedegaard:2017}, and normalized it by the fraction of deaths attributed to addicted persons ($54.6\%$) \cite{Gwira:2014} and the predicted number of deaths from our model with the age-adjusted U.S. population in 1999  ($259\times10^6$) \cite{USCensus:2018}, e.g., $A_0 = \frac{(0.546)(2749)}{(259\times10^6)(\mu^*-\mu)}=0.00136$. We then assumed  $R_0=0.1A_0$ \cite{SurgeonGeneral:2016} (fraction of population in treatment), making $S_0=0.990504$.

%
%




%
%
%

\subsection{Analysis of the Addiction-Free Equilibrium}
\label{app:AFE}

Here we derive conditions on the existence of an addiction-free equilibrium (AFE) within the system defined by Eqns.\ref{eq:S}-\ref{eq:R}.
To begin, we set each equation to zero and require that $A=0$. Eqn. \ref{eq:A} becomes $0=-(\delta+\sigma+\mu)R$, and since $\mu>0$ as a natural death rate, this implies that $R=0$ at any AFE (conversely, $R=0$ requires that either $A=0$ or $\zeta=0$, which my apply at the beginning of an epidemic). We are left with the system
\begin{align*}
    0 &= -\alpha S^\ast - \beta\xi S^\ast P^\ast + \epsilon P^\ast + \mu P^\ast\\
    0 &= \alpha S^\ast - (\epsilon + \gamma + \mu)P^\ast\\
    0 &= P^\ast(\gamma + \beta\xi S^\ast).
\end{align*}

We require that $P\neq 0$ since otherwise the only solution is $S^\ast=P^\ast=A^\ast=R^\ast=0$. Then $0 = \gamma+\beta\xi S$. Since all our parameters and dependent variables are non-negative by definition, $\gamma=0$ and either $\beta=0$ or $\xi=0$. If $\beta=0$, opioids are no longer available anywhere in our model, and so it is only natural that the addiction state dies out. If $\xi=0$, opioids are available only through the presence of current addicts (e.g. on the black market due to illicit demand) and not through currently prescribed users. In this case, we can use our assumption that $1=S+P+A+R$ to find that 
\begin{align*}
    S^\ast &= \frac{\epsilon + \mu}{\alpha + \epsilon + \mu}\hspace{1.5cm} A^\ast = 0\\
    P^\ast &= \frac{\alpha}{\alpha + \epsilon + \mu}\hspace{1.5cm} R^\ast = 0.
\end{align*}

%
%
\subsection{Calculating the Basic Reproduction Number, $R_0$}
\label{app:R0}

Assuming that $\gamma=0$ and $\xi=0$, the necessary (if $\beta\neq 0$) and sufficient conditions for the AFE to exist, Eqns. \ref{eq:A} and \ref{eq:R} reduce to
\begin{align*}
    \dot{A} &= \sigma R +  \beta SA + \nu RA - (\zeta+\mu^{*})A\\
    \dot{R} &= \zeta A - \nu RA- (\delta+\sigma+\mu) R.
\end{align*}
Using the next generation method \cite{Heffernan:2005,Diekmann:2009} with both $A$ and $R$ treated as ``infected'', we compute the matrices $F$ and $V$ as 
\[
F = \left[\begin{array}{cc}
\frac{\beta(\epsilon+\mu)}{\alpha+\epsilon+\mu} & 0\\
0 & 0\\
\end{array}\right]\ \ \ \ \text{and}\ \ \ \
V = \left[\begin{array}{cc}
\zeta+\mu^* & -\sigma\\

-\zeta & \delta + \sigma + \mu\\
\end{array}\right].
\]
Then $R_0$ is given by the spectral radius of $FV^{-1}$,
\begin{equation*}
    \begin{gathered}
    R_0 = \frac{\beta(\epsilon + \mu)}{(\alpha + \epsilon +\mu)(\mu^\ast+\zeta\Lambda)}=\frac{\beta S^*}{\mu^*+\zeta\Lambda}\\ \text{where}\ \ \ \Lambda=\frac{\delta+\mu}{\delta+\mu+\sigma},\ S^\ast = \frac{\epsilon + \mu}{\alpha + \epsilon + \mu}.
    \end{gathered}
\end{equation*}
Prevalence of opioid addicts will rise when $R_0>1$ and fall when $R_0<1$. This result is confirmed by Jacobian analysis in Appendix \ref{app:jacobian_analysis}.

%
%
\subsection{Jacobian Analysis}
\label{app:jacobian_analysis}

Before computing the Jacobian for the system \ref{eq:S}-\ref{eq:R}, note that $N$ (total number of individuals) is constant in this model and equal to 1. Therefore the system reduces to three equations,
\begin{align}
\begin{split}
    \dot{S} &= - \alpha S - \beta(1-\xi) SA - \beta\xi S(1-S-A-R) \\
    &\ \ \ + (\epsilon+\mu) (1-S-A-R) + (\delta+\mu) R+\mu^{*}A \\
    \dot{A} &= \gamma(1-S-A-R) + \sigma R +\beta(1-\xi)SA \\
    &\ \ \ + \beta\xi S(1-S-A-R) + \nu RA - (\zeta+\mu^*)A\\
    \dot{R} &= \zeta A - \nu RA- (\delta+\sigma+\mu) R,
\end{split}\label{eqn:reduced}
\end{align}
where $P= 1-S-A-R.$ The Jacobian of this equation is given in Eqn. \ref{eqn:jacobian}.

{\scriptsize{
\begin{equation}
J=\left[\begin{array}{ccc}
-\alpha-\beta(1-\xi)A + \beta\xi(S-P) - (\epsilon+\mu)
& -\beta(1-2\xi) S - (\epsilon+\mu) + \mu^* 
& \beta\xi S +\delta - \epsilon\\
\ & \ & \ \\
-\gamma + \beta(1-\xi)A + \beta\xi(P-S) & -\gamma + \beta(1-2\xi)S + \nu R -(\zeta+\mu^*)\ \ \  & -\gamma+\sigma-\beta\xi S + \nu A\\
\ & \ & \ \\
0 & \zeta - \nu R & -\nu A - (\delta+\sigma+\mu)
\end{array}\right]\label{eqn:jacobian}
\end{equation}
}}

We now focus the Jacobian evaluated at the AFE. Recall that the existence of this equilibrium requires that $\gamma=0$ and either $\beta=0$ or $\xi=0$. If $\gamma=0$ and $\beta=0$ there are no opioids left in the model, so we assume $\beta\neq0$ and that $\xi=0$. Using these parameter values and the AFE $x_0$ given by Eqns. \ref{eqn:AFE}, $J(x_0)$ is
\begin{equation*}
\left[\begin{array}{ccc}
-(\alpha+\epsilon+\mu) &\ \ \ -\frac{\beta(\epsilon+\mu)}{\alpha+\epsilon+\mu}-(\epsilon+\mu)+\mu^* 
& \delta - \epsilon \\
0 & \frac{\beta(\epsilon+\mu)}{\alpha+\epsilon+\mu} - (\zeta+\mu^*) & \sigma\\
0 & \zeta & -(\delta+\sigma+\mu)
\end{array}\right].
\end{equation*}
The characteristic polynomial is
{\footnotesize
\begin{align*}
 \nonumber p(\lambda) = -(\alpha+\epsilon+\mu+\lambda)\Bigg[&
 \lambda^2 + \Big(\zeta+\mu^*+\delta+\sigma+\mu - \frac{\beta(\epsilon+\mu)}{\alpha+\epsilon+\mu}\Big)\lambda\\
 & + \Big(\zeta+\mu* - \frac{\beta(\epsilon+\mu)}{\alpha+\epsilon+\mu}\Big)(\delta+\sigma+\mu) - \sigma\zeta
 \Bigg].
\end{align*}
}

The root $\lambda_1=-(\alpha+\epsilon+\mu)$ is always negative since we assume that $\alpha,\epsilon\geq 0$ and $\mu>0$. Thus, the AFE is stable if the remaining two roots have negative real part. Applying the Routh-Hurwitz Stability Criteria \cite{Routh:1877,Hurwitz:1895} to the degree-two polynomial $p(\lambda)$, stability requires that
\begin{align}
    \frac{\beta(\epsilon+\mu)}{\alpha+\epsilon+\mu} &< \zeta + \mu^* + \delta + \sigma + \mu\ \ \ \ \text{and}\label{eqn:RH1}\\
    \frac{\beta(\epsilon+\mu)}{\alpha+\epsilon+\mu} &< \zeta + \mu^* - \frac{\zeta\sigma}{\delta+\sigma+\mu} = \mu^* + \zeta\left(\frac{\delta+\mu}{\delta+\mu+\sigma}\right).
    \label{eqn:RH2}
\end{align}
All parameters are strictly non-negative, so Eqn. \ref{eqn:RH2} implies \ref{eqn:RH1}. Furthermore, dividing both sides of Eqn. \ref{eqn:RH2} by the right hand size, one arrives at the $R_0$ stability criterion. Numerical stability results for the AFE further confirm this analysis and are given in Appendix \ref{app:AFE_sensitivity}.

To analyze the bifurcation of this system when $R_0=1$, we follow the method described by Castillo-Chavez and Song \cite{Castillo:2004} and demonstrated by White and Comiskey \cite{White:2007} to determine the bifurcation's direction. Given the form of $R_0$, we take $\beta$ to be the bifurcation parameter and conduct our analysis around
\[
\beta^* = \frac{\mu^*+\zeta\Lambda}{S^*}.
\]
First, we define the matrix $A$ as in \cite{Castillo:2004} but, via a change of coordinates, taking $x_0$ to be the AFE and the bifurcation parameter to be $\beta$. Writing our system of differential equations as $dx/dt=f(x,\beta)$, we have
\begin{equation}
\begin{aligned}
    A &= \frac{\partial f_i}{\partial x_j}(x_0,\beta=\beta^*) = J(x_0,\beta=\beta^*) \\ 
    &= 
    \left[\begin{array}{ccc}
-(\alpha+\epsilon+\mu) &\ \ \ -\zeta\Lambda-(\epsilon+\mu) & \delta - \epsilon \\
0 & \zeta(\Lambda - 1) & \sigma\\
0 & \zeta & -(\delta+\sigma+\mu)
\end{array}\right].
\end{aligned}
\end{equation}
It is easy to check that zero is a simple eigenvalue of $A$ and that all other eigenvalues of $A$ have negative real parts. $A$ has right eigenvector $\mathbf{x}={(-S^\ast(1-\Gamma),1,\Gamma)^Tx_2}$ where $x_2$ is free, $\Gamma={\zeta/(\delta+\mu+\sigma)}$, and $\Lambda={(\delta+\mu)/(\delta+\mu+\sigma)}$. $A$ has left eigenvector $\mathbf{y} = {(0,1,1-\Lambda)y_2}$, where $y_2$ is free. The first component of $\mathbf{x}$ is negative, but since $S^*>0$ the analysis still applies \cite{Castillo:2004}. Now we let $f_k$ be the $k$th component of $f$ and set
\begin{align*}
    a &= \sum_{k,i,j=1}y_kx_ix_j\frac{\partial^2f_k}{\partial x_i\partial x_j}(x_0,\beta=\beta^*)\\
    b &= \sum_{k,i=1}y_kx_i\frac{\partial^2f_k}{\partial x_i\partial\beta}.
\end{align*}
The non-zero derivatives are
\begin{align*}
    \frac{\partial^2f_1}{\partial S\partial A} &= 
    \frac{\partial^2f_1}{\partial A\partial S} = -\beta^*\\
    \frac{\partial^2f_2}{\partial S\partial A} &= \frac{\partial^2f_2}{\partial A\partial S} = \beta^*\\
    \frac{\partial^2f_2}{\partial A\partial R} &= \frac{\partial^2f_2}{\partial R\partial A} = \nu\\
    \frac{\partial^2f_3}{\partial A\partial R} &= \frac{\partial^2f_3}{\partial R\partial A} = -\nu\\
    \frac{\partial^2f_1}{\partial A\partial\beta} &= -S^*\\
    \frac{\partial^2f_2}{\partial A\partial\beta} &= S^*
\end{align*}
and
\begin{align*}
    a &= (1)(-S^*(1+\Gamma))(1)\beta^*+(1)(1)(-S^*(1+\Gamma))\beta^*+(1)(1)\Gamma\nu\\
    &\ \ \ +(1)\Gamma(1)\nu+(1-\Lambda)(1)\Gamma(-\nu)+(1-\Lambda)\Gamma(1)(-\nu)\\
    &=-2S^*(1+\Gamma)\beta^*+2\Lambda\Gamma\nu = -2(1+\Gamma)(\mu^*+\zeta\Lambda)+2\Lambda\Gamma\nu\\
    b &= (1)(1)S^*>0.
\end{align*}
To make $a>0$, we therefore need
\begin{equation}
\Lambda\Gamma\nu > (1+\Gamma)(\mu^*+\zeta\Lambda).
\end{equation}
If this condition is satisfied, there will be a backward bifurcation at $R_0=1$. Practically speaking, this implies that when Eqn. \ref{eqn:bcond} is satisfied, a positive, stable, endemic equilibrium exists simultaneously with the stable addiction-free state, raising the possibility that additional effort beyond reducing $R_0<1$ may be required to arrive at the addiction-free state. For our estimated parameters (see Table \ref{tab:params}), a forward bifurcation occurs for all $\zeta\geq0$.

%
%

\subsection{Addiction Free Equilibrium Numerical Analysis}
\label{app:AFE_sensitivity}

\begin{figure}[htbp]
    \centering
    \includegraphics[width=0.975\linewidth]{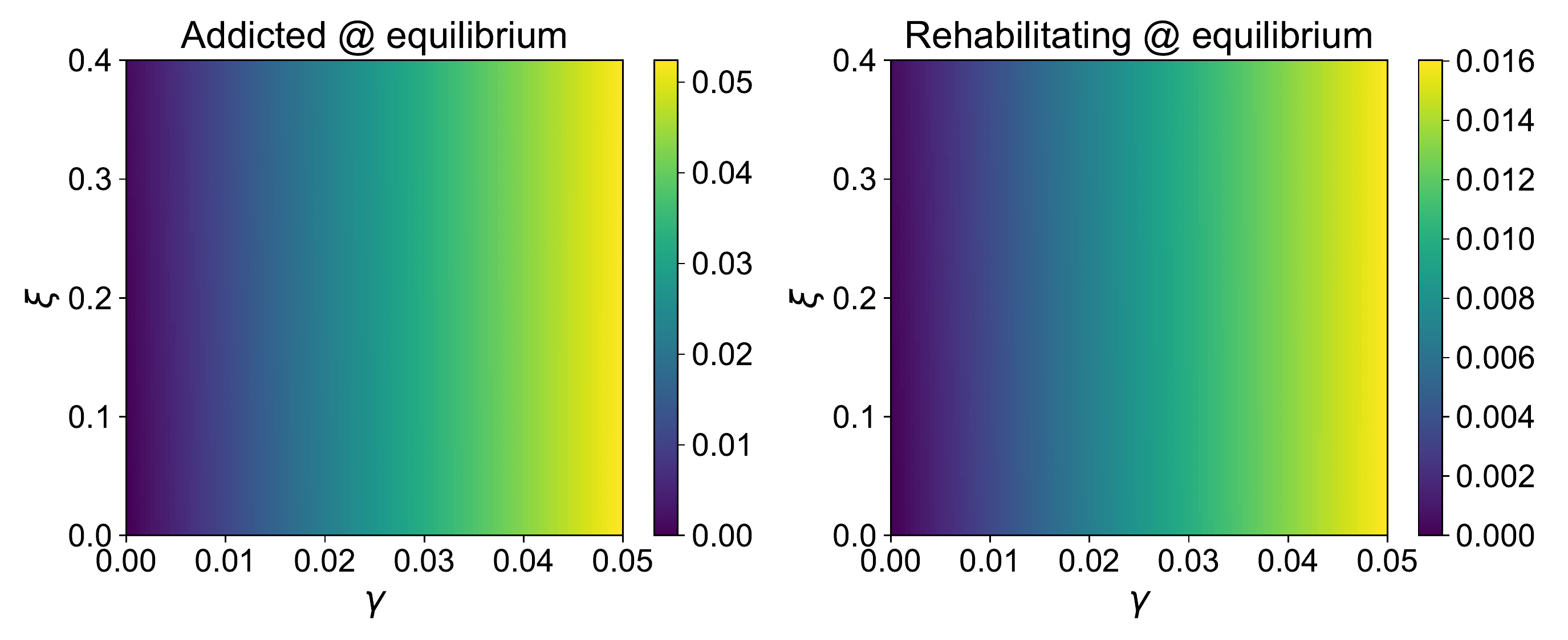}
    \caption{\textbf{Model Sensitivity to $\gamma$ and $\xi$.} Effect of moving $\gamma$ and/or $\xi$ away from zero when $R_0\approx 0.085$ with likely parameter values, $\epsilon=3$ and $\zeta=0.25$}
    \label{fig:gam_xi}
\end{figure}

To examine the sensitivity of the model's addiction free equilibrium (AFE) to its parameters, we first ran simulations to see how the AFE changes when either $\gamma$ or $\xi$ shifts away from zero. Parameter values were chosen as in Table \ref{tab:params} with $\epsilon=3$ and $\zeta=0.25$. Our results show that for our estimated parameters resulting in $R_0\approx 0.085$, shifting $\xi$ away from zero has little noticeable effect while shifting $\gamma$ away from zero strongly moves the equilibrium away from the addiction-free state (see Fig. \ref{fig:gam_xi}). This suggests that in a nearly addiction-free population, prescription-induced addiction remains far more important than securing prescriptions away from non-prescribed users. Note that in the exact case of an AFE, it is always stable when $\gamma=\xi=0$ for a parameter space centered around the other parameters listed in Table \ref{tab:params}. 

Further analysis of the model parameter space when $\gamma=\xi=0$ was conducted using the Sobol method \cite{Sobol:2001}. We chose $N=800000$ and generated $N(2D+2)$ parameter sets (where $D=9$ is the dimension of the parameter space) via Saltelli's extension of the Sobol sequence \cite{Saltelli:2002,Saltelli:2010} for a total of 16 million samples. We then ran the model to 10000 years for each set of parameters to arrive at an equilibrium. We subsequently conducted Sobol analysis \cite{Sobol:2001} on the values for $S,P,A,$ and $R$ after the final year. Initial conditions for each simulation were $S(0)=0.9435$, $P(0)=0.05$, $A(0)=0.0062$, and $R(0)=0.0003$.

\begin{figure}[htbp]
    \centering
    \includegraphics[width=\linewidth]{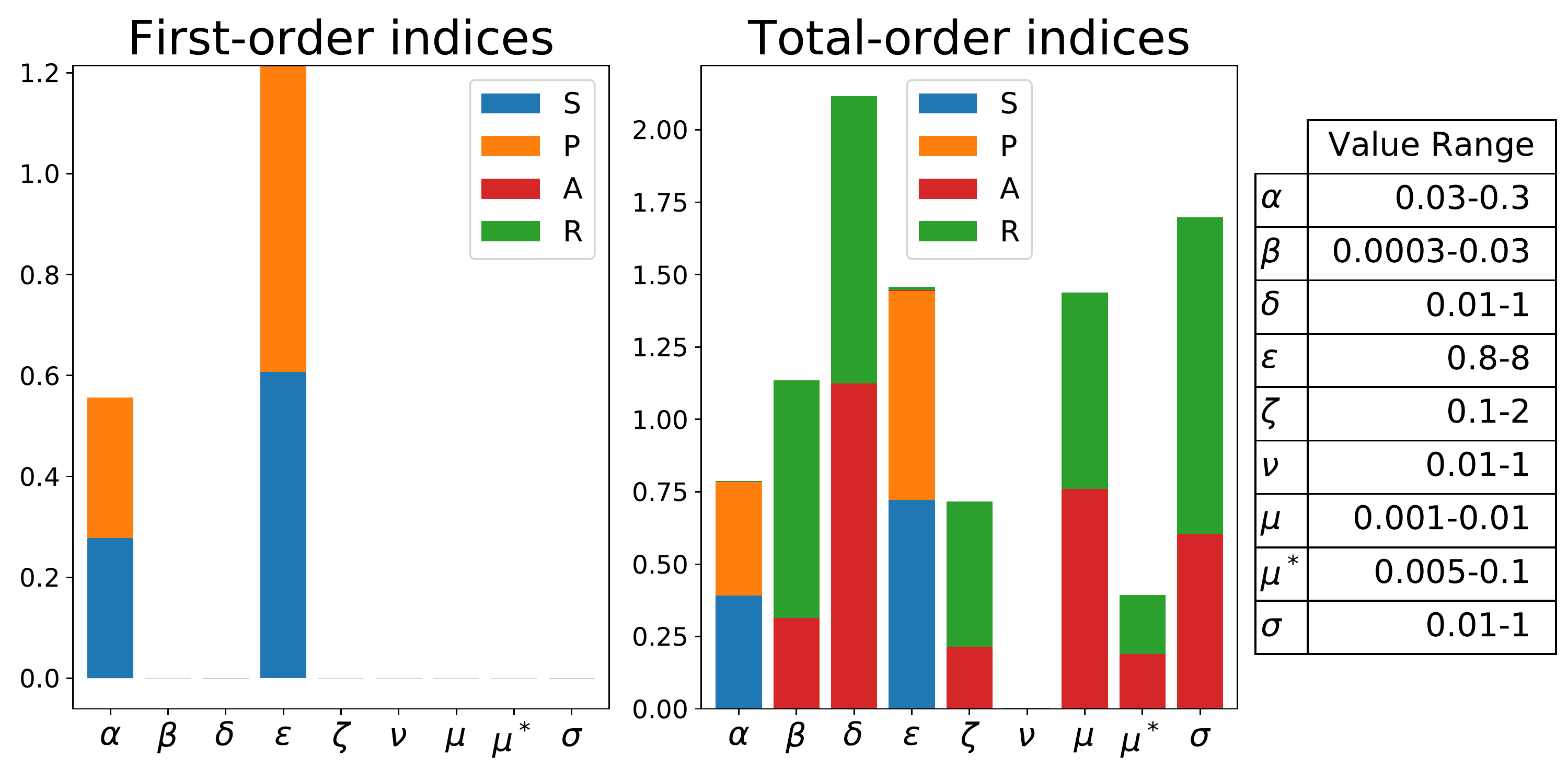}
    \caption{\textbf{Sobol sensitivity analysis for equilibrium values when $\gamma=\xi=0$} (see Fig \ref{fig:schematic} or Table \ref{tab:params} for parameter definitions). First-order indices do not take into account interactions with other parameters, while total-order indices measure sensitivity through all higher-order interactions. The parameter ranges tested here are the same as in Fig. \ref{fig:10yr_sens}}
    \label{fig:R0_sens}
\end{figure}

%
%

\subsection{Further Numerical Exploration of Parameter Space}
\label{app:numerical_exploration}


In this section we expand our parameter space exploration for $\{\epsilon,\zeta\}\in[0.8,8.0]\times[0.2,2.0]$ by examining parameter sensitivity for each of $S,P,A,$ and $R$ instead of only the addicted class. More specifically, we examine the associated effects of $\epsilon$ and $\zeta$ on the predicted populations for 10 years into the future for each of the following cases: 
\begin{enumerate}
    \item Prescription Addiction Rate ($\gamma$),
    \item Treatment Success Rate ($\delta$),
    \item Prescription Rate ($\alpha$),
    \item Method of Relapse ($\kappa$) when $\xi=0.74$
    \item Prescription Rate vs. Prescription-Induced Addiction ($\alpha$ vs. $\gamma$).
\end{enumerate}

%
%


%
%

\begin{figure}[ht]
\begin{minipage}[t]{\linewidth}
    \centering
    \includegraphics[width=0.75\linewidth]{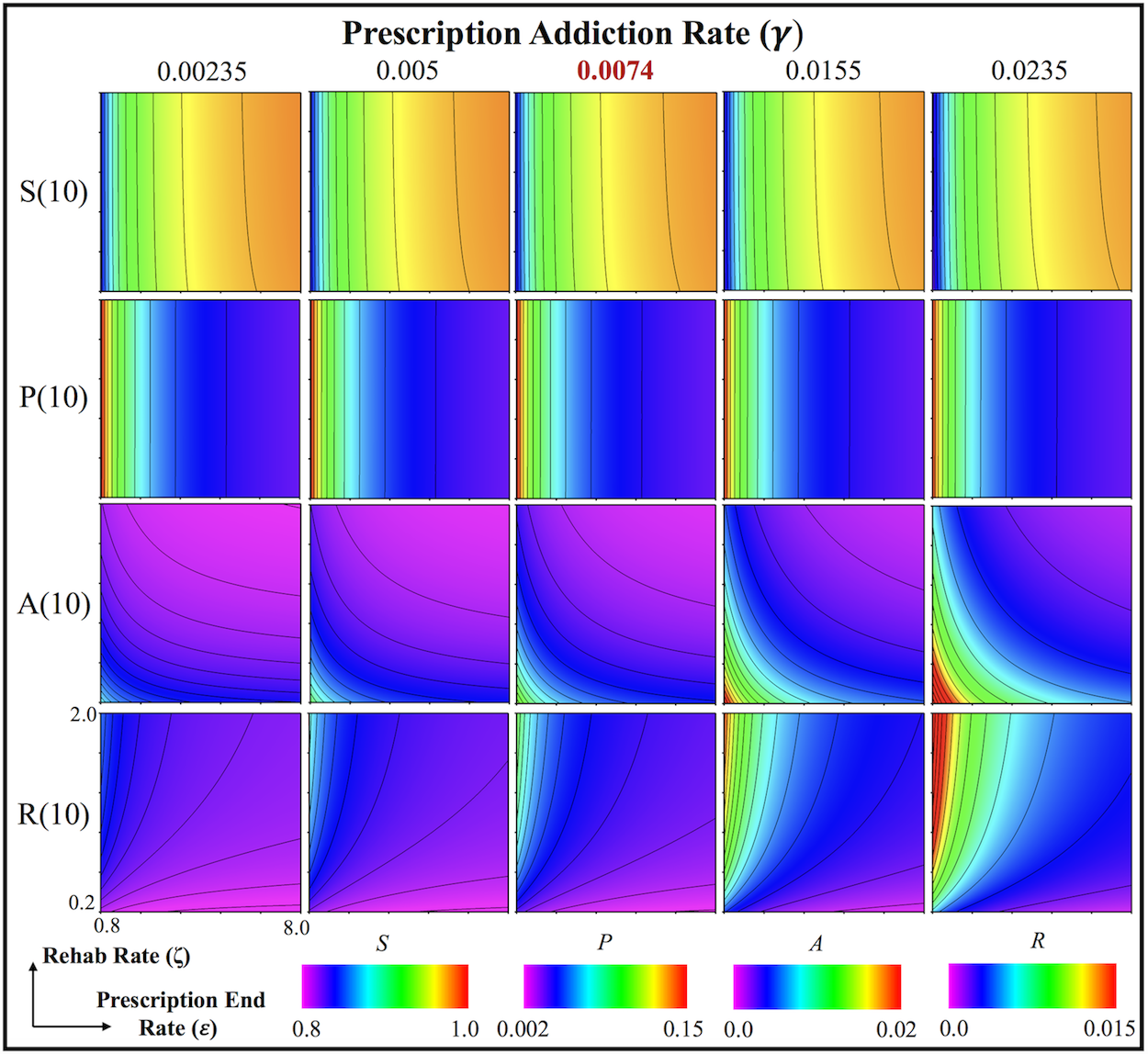}
    \caption{\textbf{Prescription Addiction Rate} Colormaps illustrating the long-term equilibrium solutions ($S^*,P^*,R^*$, and $A^*$) for prescription-end rates ($\epsilon$) and rehabilitation-start rates ($\zeta$) between $[0.8,8]$ and $[0.2,2.0]$, respectively, and for various prescription addiction rates ($\gamma$)}
    \label{fig:sims_gamma}
\end{minipage}
\begin{minipage}[t]{\linewidth}
    \centering
    \includegraphics[width=0.75\linewidth]{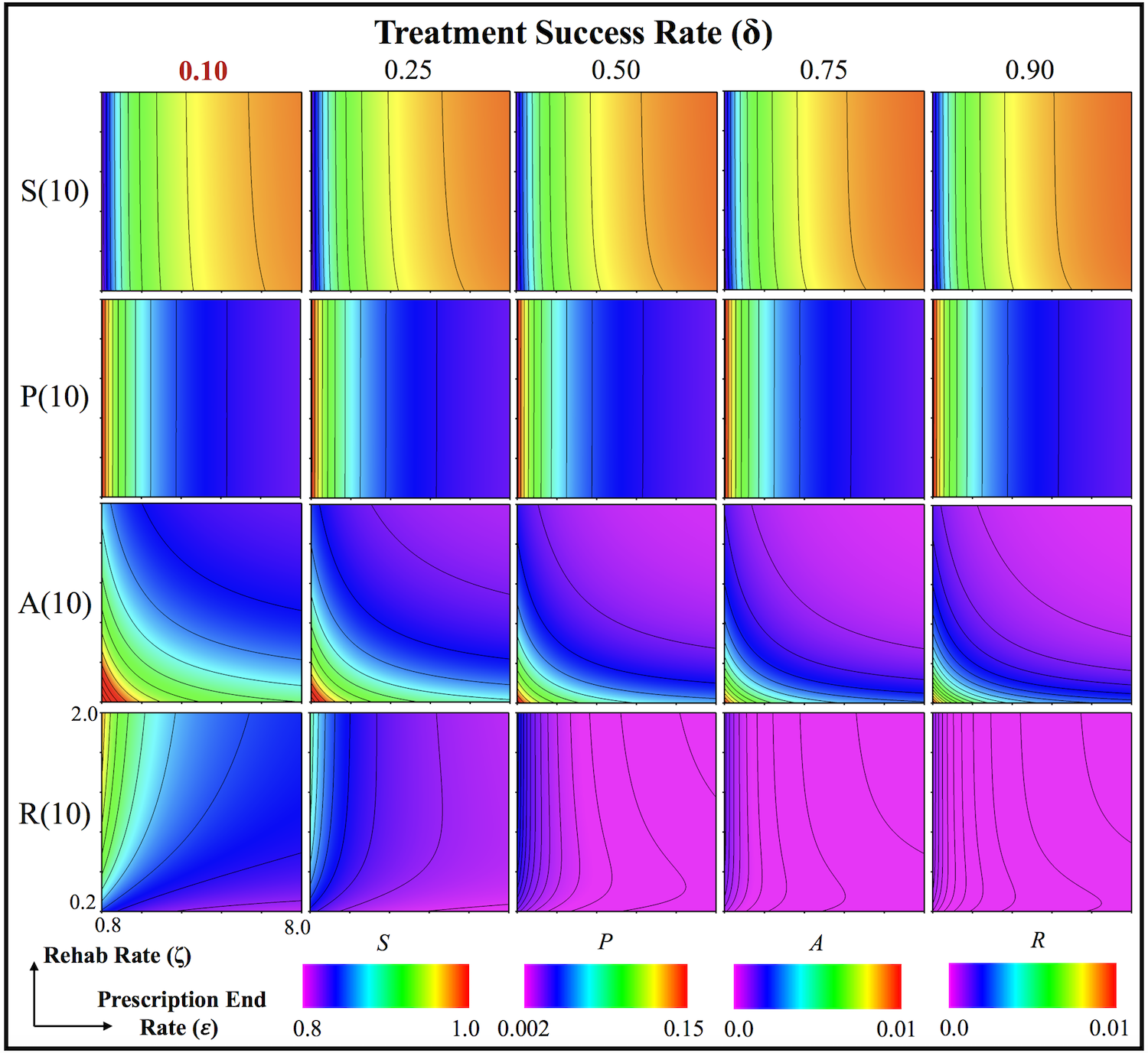}
    \caption{\textbf{Treatment Success Rate} Colormaps illustrating the long-term equilibrium solutions ($S^*,P^*,R^*$, and $A^*$) for prescription-end rates ($\epsilon$) and rehabilitation-start rates ($\zeta$) between $[0.8,8]$ and $[0.2,2.0]$, respectively, and for various treatment success rates ($\delta$)}
    \label{fig:sims_delta}
\end{minipage}
\end{figure}

As in Fig. \ref{fig:combined_results}, Fig. \ref{fig:sims_gamma} shows that as $\gamma$ increases the addicted population grows. In particular, if $\gamma$ doubles from its estimated value, there exists $(\epsilon,\zeta)$ for which $2\%$ of the population becomes addicted to opioids, which is approximately three times the number of addicts in 2016. Moreover, as $\gamma$ increases, so does the rehabilitation class. Interestingly, for values of $(\epsilon,\zeta)$ that make the addicted class roughly $2\%$ of the population, the rehabilitation class makes up approximately $1\%$. On the other hand, when the rehabilitation class composes roughly $1.5\%$ of the population, the addicted class makes up roughly the same percentage.
%
%
When $\delta$ increases the rehabilitation class population decreases near zero. The population of the addicted class decreases towards zero as well, while the populations of the susceptible class and prescribed class appear unaffected (Fig. \ref{fig:sims_delta}). 

%
%

\begin{figure}[ht]
\begin{minipage}[t]{\linewidth}
    \centering
    \includegraphics[width=0.75\linewidth]{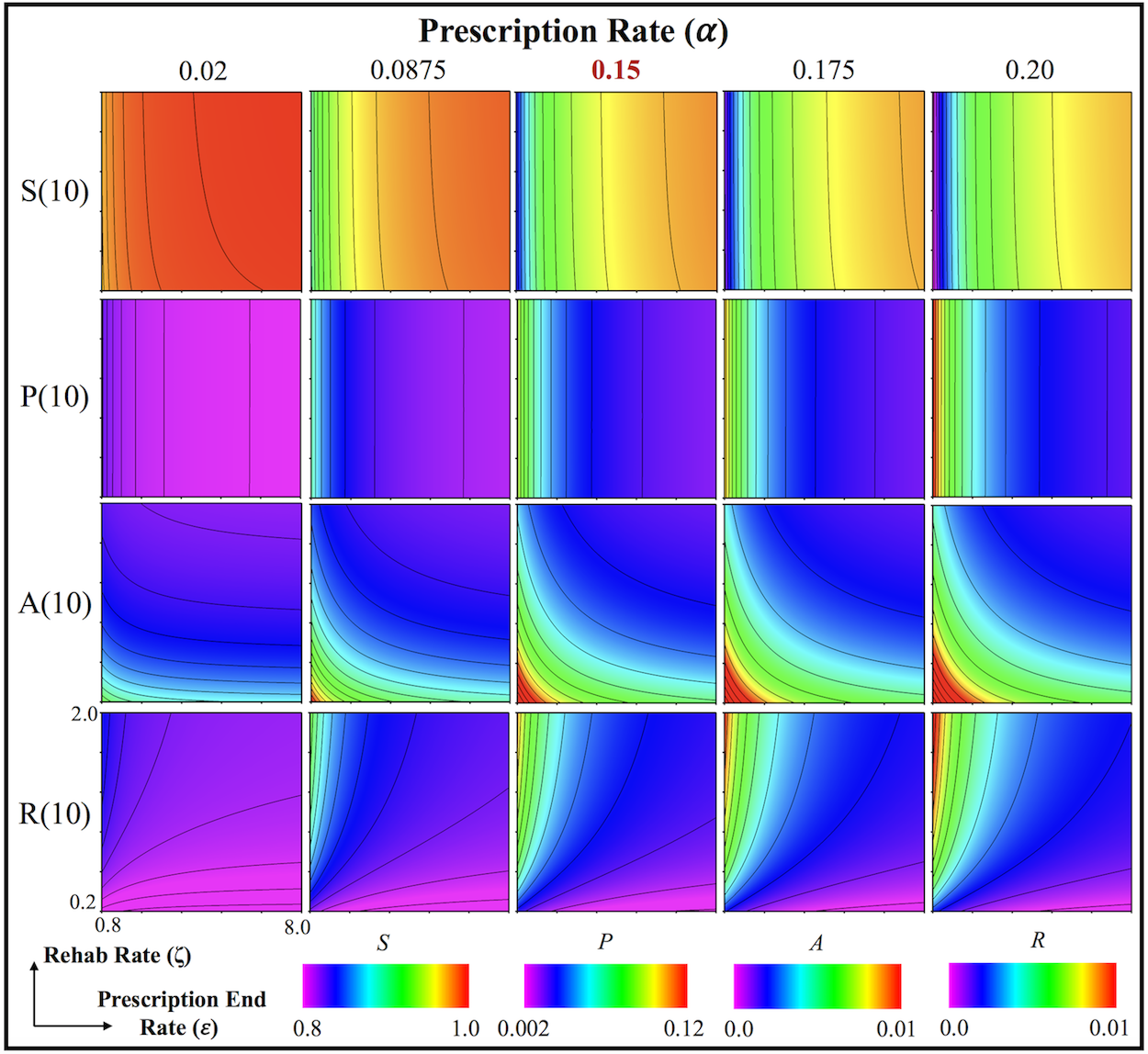}
    \caption{\textbf{Prescription Rate} Colormaps illustrating the long-term equilibrium solutions ($S^*,P^*,R^*$, and $A^*$) for prescription-end rates ($\epsilon$) and rehabilitation-start rates ($\zeta$) between $[0.8,8]$ and $[0.2,2.0]$, respectively, and for various prescription rates ($\alpha$)}
    \label{fig:sims_alpha}
\end{minipage}
\begin{minipage}[t]{\linewidth}
    \centering
    \includegraphics[width=0.75\linewidth]{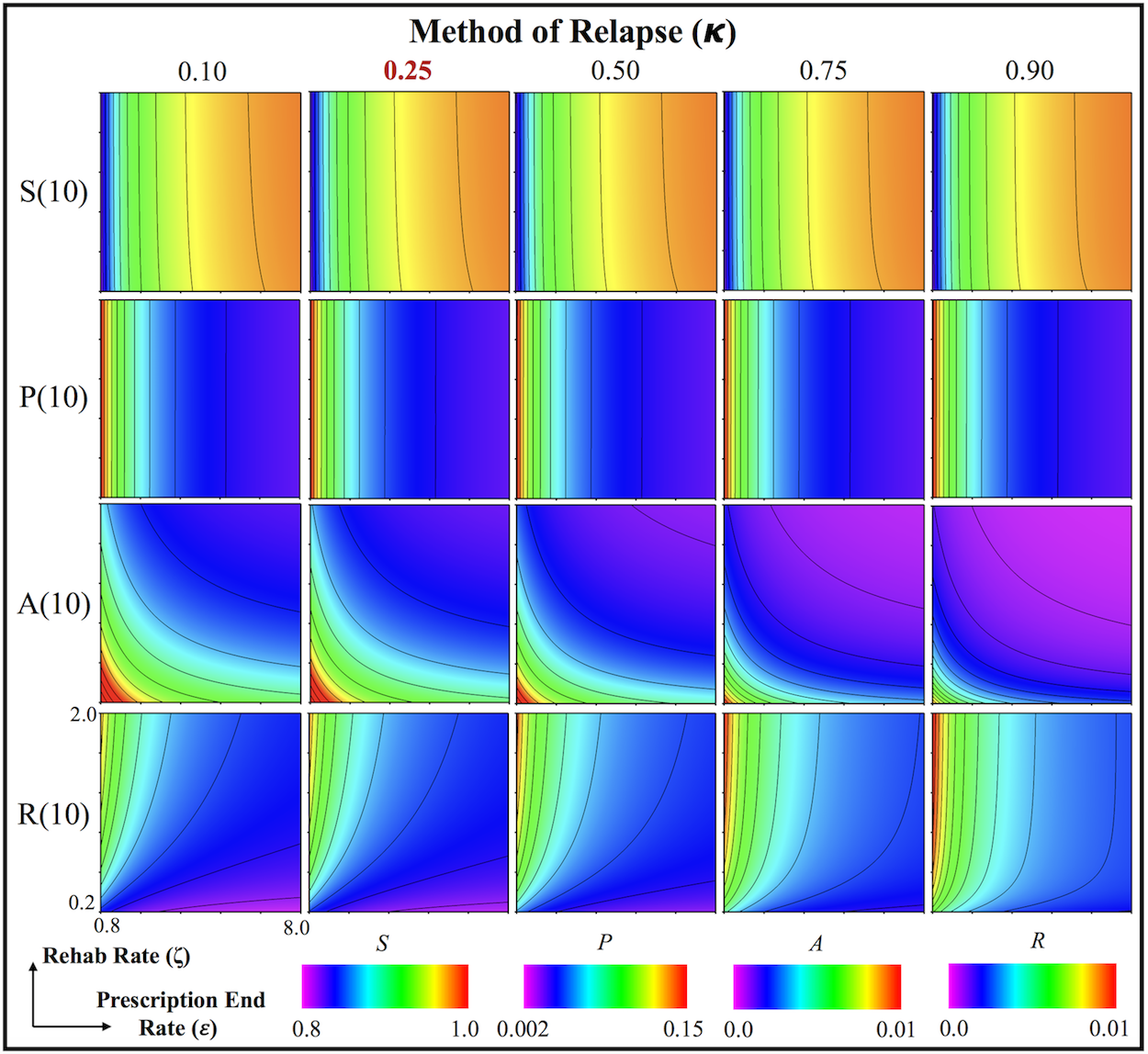}
    \caption{\textbf{Relapse Due to Overall Illicit Usage and Availability.} Colormaps illustrating the long-term equilibrium solutions ($S^*,P^*,R^*$, and $A^*$) for prescription-end rates ($\epsilon$) and rehabilitation-start rates ($\zeta$) between $[0.8,8]$ and $[0.2,2.0]$, respectively, and for various ratios of relapse back into addiction due to overall illicit usage and availability over the total relapse rate ($\kappa$)}
    \label{fig:model_relapse_methods}
\end{minipage}
\end{figure}

Fig. \ref{fig:sims_alpha} shows that if the prescription rate $\alpha$ is small enough, the entire population almost remains in the susceptible class. However, for certain values of $(\epsilon,\zeta)$ roughly $0.5\%$ of the population can still remain in the addicted population. Moreover, for all cases of $\alpha$ and small $\zeta$,  the rehabilitation class' population remains near zero for almost all values of $\epsilon$. 
%
%

Setting $\delta=0.1$, we let $\kappa=\nu/(1-\delta)$ be the fraction of relapse back into addiction attributable to total illicit usage and availability. $\sigma$ is then defined as $\sigma = 1-\delta-\nu$. Using the parameters in Table \ref{tab:params}, we explored the system's sensitivity to $\kappa$. The simulation data is presented in Fig. \ref{fig:model_relapse_methods}. Qualitatively, there are only subtle differences in the dynamics, but for larger values of $\epsilon$ and $\zeta$, lowering $\sigma$ relative to $\nu$ is more effective at diminishing the addicted class.

\begin{figure}
    \centering
    \includegraphics[width=0.925\linewidth]{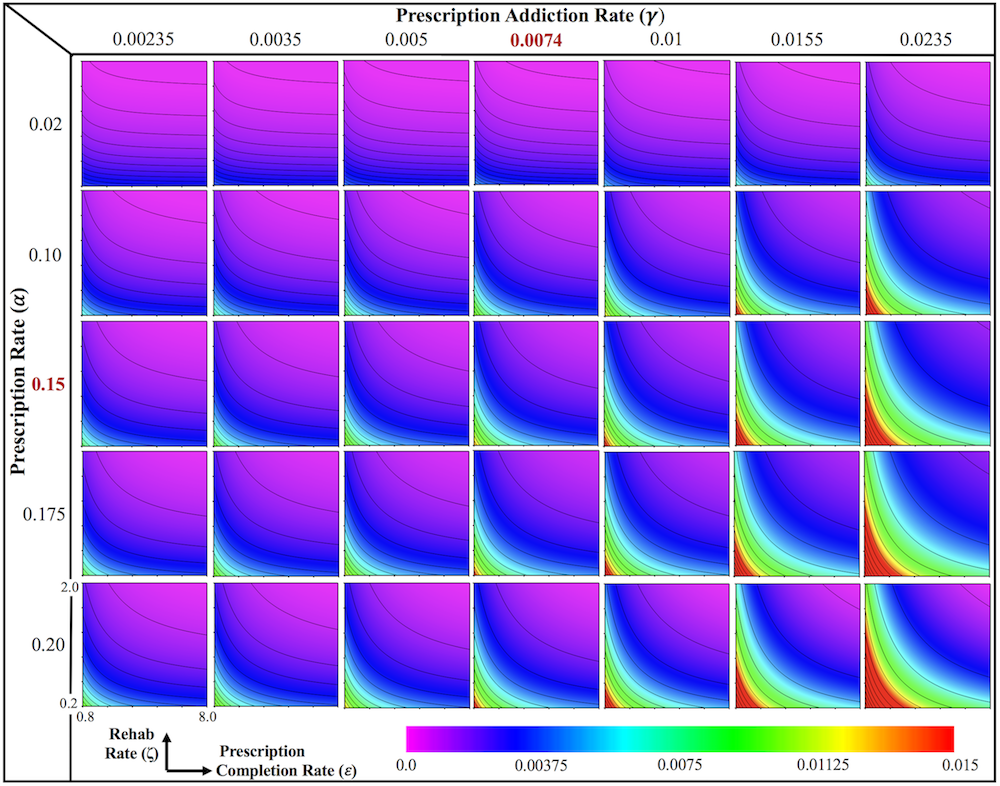}
    \caption{\textbf{Prescription-Induced Addiction vs. Prescription Completion.} Colormaps illustrating the long-term equilibrium solutions ($S^*,P^*,R^*$, and $A^*$) for prescription rates ($\alpha$) and rehabilitation rates ($\zeta$) between 0 and 1 and for various rates of prescription-induced addiction ($\gamma$) and rates of finishing prescriptions ($\epsilon$)}
    \label{fig:gamma_vs_alpha}
\end{figure}

Finally, we explore the relationship between prescription-induced addiction ($\gamma$) and completing the prescription and heading back into the susceptible class ($\epsilon$). Situations in which these two parameters do not add to one could be used to model long or short-term opioid prescription use. The data is presented in Fig. \ref{fig:gamma_vs_alpha}. It is clear that a decrease in $\epsilon$ corresponds to an increase in the number of addicts as might be expected for more chronic opioid prescription use. For large $\gamma$ those differences are more subtle, as increasing $\gamma$ leads to a profound escalation in the addicted population regardless of $\epsilon$.

\clearpage

%
%

\bibliographystyle{spmpsci}      

\bibliography{heroin_heart}

\end{document}